\documentclass[runningheads,a4paper]{llncs}
\usepackage{amssymb}
\usepackage{graphicx}

\usepackage{url}
\urldef{\mailsc}\path|{federico, miksch}@ifs.tuwien.ac.at|    
\newcommand{\keywords}[1]{\par\addvspace\baselineskip
\noindent\keywordname\enspace\ignorespaces#1}

\usepackage{cite}
\usepackage{paralist}
\usepackage[inline]{enumitem}
\usepackage{placeins}
\usepackage{tabularx}
\usepackage{subcaption}
\usepackage{xcolor}
\usepackage{amssymb}

\definecolor{myred}{RGB}{228, 26, 28}
\definecolor{myblue}{RGB}{55, 126, 184}
\definecolor{mygreen}{RGB}{77, 174, 74}
\definecolor{mypurple}{RGB}{152, 78, 163}

\makeatletter 
\let\@tabclassz =\@classz 
\let\@tabclassiv =\@classiv 
\makeatother 

\DeclareCaptionSubType*{subfigure}
\begin{document}

\mainmatter  

\title{Evaluation of two interaction techniques for visualization of dynamic graphs}

\titlerunning{Evaluation of two interaction techniques for visualization of dynamic graphs}

%
%
\author{Paolo Federico%
\and Silvia Miksch}
\authorrunning{P. Federico and S. Miksch}

\institute{Institute of Software Technology and Interactive Systems, TU Wien,\\
Favoritenstrasse 9-11, 1040, Vienna, Austria\\
\mailsc\\
}

%
%

\toctitle{Evaluation of two interaction techniques for visualization of dynamic graphs}
\tocauthor{P. Federico and S. Miksch}
\maketitle

\begin{abstract}
Several techniques for visualization of dynamic graphs are based on different spatial arrangements of a temporal sequence of node-link diagrams. Many studies in the literature have investigated the importance of maintaining the user's mental map across this temporal sequence, but usually each layout is considered as a static graph drawing and the effect of user interaction is disregarded. We conducted a task-based controlled experiment to assess the effectiveness of two basic interaction techniques: the adjustment of the layout stability and the highlighting of adjacent nodes and edges. We found that generally both interaction techniques increase accuracy, sometimes at the cost of longer completion times, and that the highlighting outclasses the stability adjustment for many tasks except the most complex ones.
\end{abstract}

\section{Introduction}

Dynamic graphs can be used to model different complex real-word phenomena and, therefore, are collected and analysed in various disciplines. Visualization is an indispensable mean to make sense of this type of time-oriented data and gain valuable insights about the phenomena they represent. In recent years, the research about visualization of dynamic graphs has seen a rapid growth, with many novel approaches, techniques, and systems, as surveyed by recent reviews. Likewise, many evaluation studies have investigated those visualizations, to understand which are the key design factors, how they are perceived by users, and how they can support users in analysing data and solving their tasks. The evaluation of dynamic graph visualization has focused mainly on two aspects: comparing animation versus static views, and assessing the importance of the mental map preservation. 
These studies have often found conflicting results, or a high variability of results depending on different tasks. The use of interaction, in order to control the amount of mental map preservation, or to switch from animation to static views, has been proposed as a means to increase the applicability of a given visualization to diverse tasks. Nevertheless, few evaluation studies have focused on the role of interaction in dynamic graph visualization: usually static views are considered as noninteractive, while for animated views the most contemplated interaction is the playback control. 
To fill this gap, we focused on the mental map preservation and its interactive control, which we empirically evaluated in comparison with another common interaction such as highlighting. 
Thus, the contribution of this paper is a controlled task-based experiment to quantitatively evaluate two interaction techniques for dynamic graph visualization, namely the interactive control of the mental map and the interactive highlighting of adjacent nodes and links. In the following, we review the related work; list the hypotheses, the design, and the settings of our study; present the results and discuss their implications.

\section{Related work}
Herman et al.~\cite{Herman2000} provide an early survey on graphs in information visualization, focusing on layout algorithms for both the general case and special subclasses (e.g., planar graphs, trees) as well as on techniques for interactive navigation, in particular focus+context and clustering. The state of the art report by von Landesberger et al.~\cite{vonLandesberger2011}  offers an updated and extensive review of the field; it has a particular focus on issues and solutions for large scale graphs, but contains sections on dynamic graphs as well as interactions. Kerracher et al.~\cite{Kerracher2014} explore, and outline a structure of, the design space of dynamic graph visualization. Archambault et al.~\cite{Archambault2014} also review the field, discussing in particular multivariate and temporal aspects of networks. 
A comprehensive survey on visualizing dynamic graphs is found in Beck et al.~\cite{Beck2014}.


\paragraph{\textbf{The layout stability}}
\label{sec:mentalmap}
Many existing algorithms for drawing dynamic graphs ensure the layout stability in order to preserve the user's mental map of the graph~\cite{Misue1995}. This stability can be seen as an additional aesthetic criterion for dynamic graphs, prescribing that the placement of nodes should change as little as possible~\cite{Coleman1996}. The utility of this dynamic aesthetics has been highly disputed in literature and several evaluations have been conducted, from both the algorithmic and the perceptual perspective. 
As for the algorithmic evaluation, Brandes \& Mader~\cite{Brandes2012} compare three approaches: \textit{aggregation} (fixed nodes positions are obtained from the layout of an aggregate of all graphs in the sequence, achieving maximum stability), \textit{anchoring} (nodes are attracted by reference positions), and \textit{linking} (nodes are attracted by instances of themselves in adjacent time slices of the sequence). Their results suggest that the generally preferable approach is linking, that is also the most computationally demanding; a faster alternative is anchoring to an aggregate layout initialized with the previous one in the sequence. 
Many user studies have been performed to empirically assess the importance of mental map preservation for readability, memorability, or interpretability of dynamic graphs. Archambault \& Purchase~\cite{Archambault2013} review many of them. In an early study about readability of direct acyclic graphs (DAGs), Purchase et al.~\cite{Purchase2007} found that the layout stability is beneficial for several tasks. Conversely, a similar study about readability of DAGs by Zaman et al.~\cite{Zaman2011} found no significant effect of the layout stability. Purchase \& Samra~\cite{Purchase2008} tested several interpretation tasks for directed graphs and found that extremes (no stability or maximum stability) are better than a medium stability. Conversely, Saffrey \& Purchase~\cite{Saffrey2008}, by investigating  readability and interpretability of directed graphs, found that the layout stability does not provide any advantage and can be even harmful for certain tasks.  While all the evaluations mentioned so far were conducted on timeline based visualization, Ghani et al.~\cite{Ghani2012} studied the effects of layout stability in readability of animated node-link diagrams, finding that a fixed layout (maximum stability) outperforms a force-directed layout with no stability. Analogously, by studying the memorability of animated node-link diagrams, Archambault \& Purchase~\cite{Archambault2012} found that  maximum layout stability  was the best condition. 

\paragraph{\textbf{User interaction in dynamic graph visualization}}
User interaction is, by definition, a crucial aspect of Information Visualization~\cite[page 6]{Card1999}. Various motivated calls have been issued to establish a ``Science of Interaction'' to complement Information Visualization~\cite{Pike2009}. Yi et al.\cite{Yi2007} propose a taxonomy of interaction in Information Visualization based on the notion of user intent; Lam~\cite{Lam2008} introduces a theoretical framework to understand and possibly reduce the costs of interaction. Nevertheless, the importance of user interaction in dynamic graph visualization is generally underestimated in literature~\cite{Beck2014}; timeline approaches are generally considered as sequences of static (i.e., non interactive) drawings, while the most discussed interaction for animation approaches deals with animation control (e.g., play/pause, or time seeker). 
%
Wybrow et al.~\cite{Wybrow2014} review interaction techniques for multivariate graphs and propose a classification based on the Information Visualization Reference Model~\cite{Card1999}, distinguishing among view level, visual-representation level, and data level. 
Notable examples include a technique for selecting and manipulating subgraphs~\cite{McGuffin2009} or a network-aware navigation integrating ``Link sliding'' (guided panning when dragging along a visible link) and ``Bring \& Go'' (bringing adjacent nodes nearby when pointing to a node), with the latter having the best performance~\cite{Moscovich2009}.   
%
Another example of evaluating interaction in dynamic graph visualization is provided  by Rey \& Diehl~\cite{Rey2010}, who investigate the effects of two interaction techniques for animated visualization: interactive control of the animation speed and a tooltip showing details on demand. They found that the speed control does not provide a significant benefit, and the tooltip is outperformed by a visualization having labels always visible. 

Given the high variability in the importance of the mental map (depending on tasks, user preferences, and possibly other factors), some scholars have proposed an interactive control of the layout stability, to let the user fine-tune it~\cite{Bach2014},\cite{Federico2011}. According to the taxonomy of interaction by Yi et al.~\cite{Yi2007}, intaractive control of stability can be understood as a \textit{Reconfigure} interaction, corresponding to the user's intent: ``Show me a different spatial arrangement''. It falls into the class of user-controlled adjustments of the layout, which are very common visual-level interactions for graphs~\cite{Wybrow2014}. Bach et al.~\cite{Bach2014} evaluated this stability slider in the context of a specific technique (GraphDiaries) featuring staged animations of node-link diagrams, but they did not consider it as an independent factor in the study design. Smuc et al.~\cite{Smuc2014} also evaluated a graph visualization featuring a stability slider, but without a specific focus on it. 

Layout stability has been also described as a form of spatial highlighting, where position is used to identify different instances of the same node over time~\cite{Archambault2014b}. 
\textit{Highlighting}, in the stricter sense, is a \textit{brushing} interaction technique, originally developed for scatter plots\cite{Becker1987}, and then extended and applied to other visualization techniques. Brushing is a ``change in the encoding of one or more items essentially immediately following, and in response to, an interaction with another item''~\cite[p. 235]{Spence2007}. 
In particular, in the case of highlighting, the change may affect hue, brightness, or color. Brushing operates within a view or across multiple views; in the latter case, the interaction technique is better known as \textit{linking and brushing}~\cite{Keim2002}. 
Highlighting makes some information stand out from other information; it effectively exploits pre-attentive processing~\cite{Ware2004}, which is the human capability to process visual information prior to, or in the early stage of, focusing conscious attention. Linking and brushing techniques support two user's intents: \textit{Select}, i.e., ``Mark something as interesting'' and \textit{Connect}, i.e., ``Show me related items'', according to the interaction taxonomy by Yi et al.~\cite{Yi2007}.  
In the context of graph visualization, highlighting of adjacent nodes upon selection of a certain node (for example, by mouse hovering) is a common interaction technique, also known as \textit{connectivity highlighting}~\cite{Heer2005}.  
An experiment by Ware \& Bobrow has shown that interactive highlighting can efficiently support visual queries on graphs~\cite{Ware2005}. In the case of timeline visualization of dynamic graphs, the highlighting technique can be extended in order to fulfil the need of visually linking and synchronizing multiple instances of the same graph entities in different time slices~\cite{Beck2014}, by considering adjacency not only across the graph structure, but also along the time dimension. 


\paragraph{\textbf{Evaluation of other aspects in graph visualization}}
Besides the importance of preserving the mental map in node-link diagrams, another issue which attracts the interest of scholars is the comparison between animation approaches and timeline approaches, the latter being usually based on small multiples in a juxtaposition arrangement. The controlled experiment by Farrugia and Quigley~\cite{Farrugia2011} found that static drawings outperform animation in terms of task completion time. Archambault et al.~\cite{Archambault2011}, in an analogous user study, found that small multiples are generally faster, but more error-prone for certain tasks; moreover, mental map preservation has little influence on both response time and error rate. Boyandin et al.~\cite{Boyandin2012} also conducted a comparative evaluation of animation versus small multiples in the context of flow maps. They found that  with animation there were more findings of changes in adjacent time slices, where with small multiples there were more findings about longer time periods. Moreover, they suggest that switching from one view to the other might lead to an increase in the numbers of findings; we see this consideration in analogy with the mental map case, where the introduction of a stability slider might allow the user to adapt the layout to a particular task and possibly increase the overall visualization effectiveness.

\paragraph{\textbf{Task taxonomies}}
A profound understanding of analytical tasks is a necessary prerequisite to design novel visualization techniques as well as evaluate existing ones. 
Ahn et al.~\cite{Ahn2014} propose a task taxonomy for dynamic graphs along three different axes: graph entities%
, temporal features%
, and properties (structural and domain-specific). 
According to Bach et al.~\cite{Bach2014}, each task can be understood as a question containing references to two dimensions and requiring an answer in the third one. In this way, they distinguish between 
topological tasks%
, temporal tasks%
, and behavioural tasks.
%
Archambault and Purchase~\cite{Archambault2014} structure their taxonomy along two dimension, mostly aiming at assessing the importance of the mental map. They distinguish between local and global tasks%
, and between distinguishable and undistinguishable tasks (depending on whether graph entities need to be distinguished from each other or can be aggregated).
A task taxonomy for multivariate networks can be found in~\cite{Pretorius2014}.  

\paragraph{\textbf{Open challenges}}
Summarizing our review of related work on visualization of dynamic graphs, we can observe that many techniques have been designed and evaluated, but the research community lacks final and well-established results about highly disputed issues, such as the importance of the layout stability, or the comparison between animation and timeline approaches. In both cases, it has been suggested that enabling users to interactively switch between different views might broaden the number of tasks that they can efficiently solve. 
Hence, there is a commonly recognised need of understanding the role of interaction in the context of dynamic graph visualization, but only few studies have specifically focused on the evaluation of interaction techniques.

\section{Our evaluation}
\begin{figure*}[bth]
 \centering
 \includegraphics[width=\textwidth]{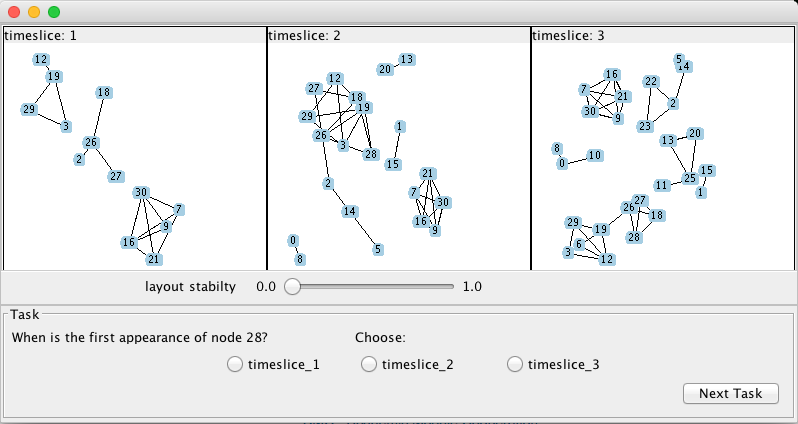}
 \caption{The remote evaluation software displays stimuli, provides instructions, and measures time and error.}
 \label{screenshot}
\end{figure*}

Addressing the aforementioned need, we performed a user study to explore interaction in the context of dynamic graphs visualization. In particular, we considered a timeline visualization with the juxtaposition approach, where several node-link diagrams are arranged along a horizontal timeline (Figure \ref{screenshot}). 
We evaluated two interaction techniques. The first one is the interactive control of the layout stability, which is executed by the means of a slider control (thus, for the sake of brevity, we will refer to it as the \textit{Slider}). The second interaction is the highlighting of adjacent nodes, adapted for dynamic graphs (in the following, \textit{Highlighting}). In this section we detail the  study design, the stimuli, the tasks, and the settings of our empirical experiment. 

\paragraph{\textbf{Study design}} 
%

We designed our user study as a quantitative controlled task-based evaluation, with two observed dependent variables: time and error. 
 We considered two factors, i.e. independent variables: the presence of the \textit{Slider} interaction ($2$ levels: off/on), and the presence of the \textit{Highlighting} interaction ($2$ levels: off/on). 
In other words, we considered four different interfaces: no-interaction, only \textit{Slider}, only \textit{Highlighting}, and both interactions. We chose this design in order to compare the two interaction with each other and with the non-interactive baseline, and also to assess how the two interactions work together. 

We tested $6$ \textit{task} types. The full factorial design led to a total amount of $N = Task  \times Slider \times Highlighting=6 \times 2 \times 2 = 24$ conditions. To mitigate the effects of personal skills and preferences, we chose a within-subject design; each subject tested $24$ conditions, by solving a different task for each of the six task type and for each of the four interfaces. In order to lower the cognitive effort of switching between different interfaces, we grouped conditions by interface. To mitigate the effects of learning and fatigue, we used a Latin square arrangement of the interfaces and we randomized the order of the tasks within each interface. Moreover, we randomized the initial slider position.

\begin{figure}[hbt]
 \centering
 \includegraphics[width=\columnwidth]{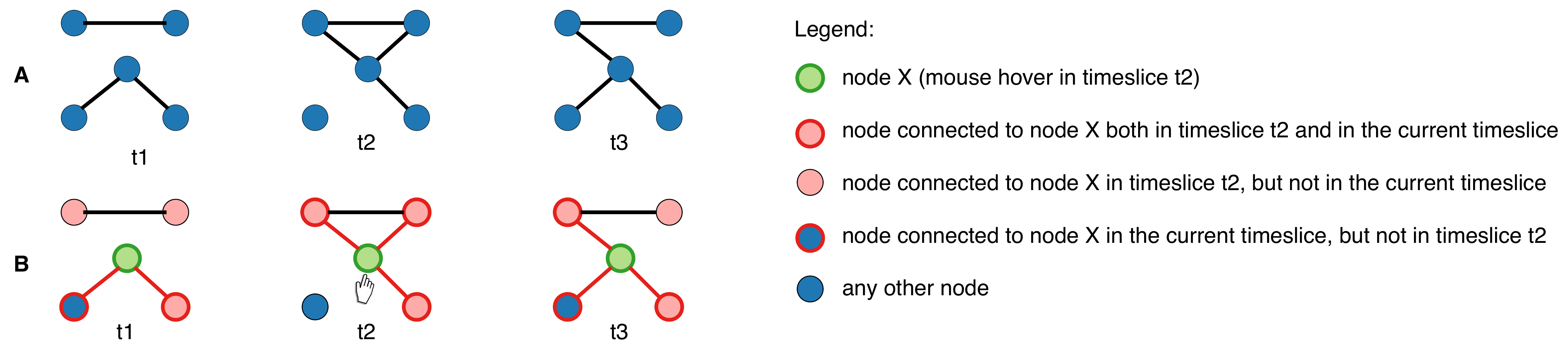}
 \caption{A: a dynamic graph over three time slices; B: the same graph highlighted on a mouse-over event.}
 \label{Highlighting}
\end{figure}

\paragraph{\textbf{Stimuli}}
We selected as baseline a spring-embedder layout as implemented in the Prefuse visualization toolkit~\cite{Heer2005a}(Figure \ref{screenshot}). According to the \textit{linking} approach~\cite{Brandes2012}, we added inter-time links to the graph in order to ensure layout stability. The \textit{Slider} controls the amount of stability by interactively changing the relaxed lengths of inter-time springs. 
%
%
In the implementation of the \textit{Highlighting} technique used for our experiment, when the user moves the mouse pointer over a node, a different combination of fill and stroke colors is used to highlight each different type of ``adjacent'' graph item, as shown in Figure~\ref{Highlighting}.

For the experiment we used real-world datasets: the dynamic graphs of social relationship between university freshmen collected by van Duijn, consisting of $38$ nodes and $5$ time points, and the one collected by van de Bunt, consisting of $49$ nodes and $7$ time points~\cite{VanDeBunt1999}. Through a threshold mechanism we derived two dynamic unweighted (i.e., binary) graphs from the original dynamic weighted graphs. Each task involved only subsets of $3$ time slices.

\paragraph{\textbf{Tasks}}
We selected six different types of tasks (Table \ref{tab:tasks}). 
As a criterion for the selection of the tasks, we considered existing studies on the importance of layout stability and tried to elicit a set of similar tasks, in order to have comparable results. Furthermore, we considered the task taxonomy by Ahn et al.~\cite{Ahn2014} in order to have a meaningful and representative set of tasks along its three axes. 
As for the graph entities, we included tasks referring to all the levels: the \textit{entity} level (nodes, links), the \textit{group} level (paths, components), and the \textit{graph} level (the entire graph). 
As for the properties, we disregarded tasks referring to \textit{domain properties} and only considered tasks referring to \textit{structural properties}, which are specific aspects for graphs. As for the temporal features, we included tasks referring to \textit{individual events} and \textit{contraction \& growth}, scoping out more complex tasks, which can be investigated in a follow-up study.   
In order to better describe the nature of our tasks and to enable a better interpretation of results, we also categorized our tasks according to other existing taxonomies \cite{Archambault2013,Bach2014}, as shown in Table \ref{tab:tasks}. 

\begin{table*}[bht]
 \caption{Task types, examples, and classifications}
 \label{tab:tasks}
 \begin{center}
   \begin{tabularx}{\textwidth}{c|l|l|l|l}
    Task  & Description & by~\cite{Ahn2014} & \cite{Archambault2013}  & \cite{Bach2014} \\
   \hline
1.NO & Node Occurence & Event/Node & Local/Distinguishable & When \\ 

 & \multicolumn{4}{l}{e.g., \textit{When is the first appearance of node 27?}} \\

   \hline

2.LO & Link Occurence & Event/Link & Local/Distinguishable & When \\

 & \multicolumn{4}{l}{e.g., \textit{When is the last appearance of link 6--9?}} \\

   \hline

3.ND & Node Degree & Event/Group & Local/Indistinguishable & When  \\

 & \multicolumn{4}{l}{e.g., \textit{When does the smallest degree (number of connections) of node 10 occur?}} \\

   \hline

4.SP & Shortest Path & Event/Group  & Local/Distinguishable & When \\

 & \multicolumn{4}{l}{e.g., \textit{When does the largest geodesic distance between node 7 and node 9 occur?}} \\

   \hline

5.CC & Connected Components & Growth/Graph & Global/Indistinguishable & What  \\

 & \multicolumn{4}{l}{e.g., \textit{Is the number of connected components increasing, decreasing, or stable?}} \\

   \hline

6.AL & All Links & Growth/Graph & Global/Indistinguishable & What  \\

 & \multicolumn{4}{l}{e.g., \textit{Is the total number of edges increasing, decreasing, or stable?}} \\

   \end{tabularx}
 \end{center}
\end{table*}
  

\paragraph{\textbf{Subjects' pool and study settings}}
We conducted the experiment remotely by using the Evalbench toolkit \cite{Aigner2013} (Figure~\ref{screenshot}). 
In order to assess the technical setup, the estimated overall length of the evaluation session, and the understandability of textual descriptions of our tasks, we performed two pilot tests with direct observation of subjects, and then we implemented small adjustments before the main remote study.  
For the main study we recruited 64 volunteer subjects among undergraduate students at the fifth semester of a bachelor programme in Visual Computing. All the subjects had normal or corrected vision. 
%
%
Right after the recruiting, we instructed the subjects with a $15$ minute briefing, describing the visualization and the interactions to be evaluated, and recalling the necessary concepts from graph theory (e.g., the notion of geodesic distance as shortest path, or the notion of connected components). The subjects were instructed to be fast and accurate in solving the tasks, without assigning any priority between speed and accuracy. The evaluation software included a training session for each of the four interfaces. During the training sessions, the software does not collect data; it shows the correct answer after completion of each task and allows repetitions until the subject feels confident of having understood the task types and the interface. The test, including the training sessions, had an average duration of $20$ minutes. 

\paragraph{\textbf{Hypotheses}}
We designed our experiment to test three hypotheses:
\begin{enumerate*}[label={\Alph*)},font={\bfseries}]
\item the \textit{Slider} reduces error rates at the cost of longer completion times, in comparison with the non-interactive interface;
\item the \textit{Highlighting} reduces error rates at the cost of longer completion times, in comparison with the non-interactive interface;
\item the \textit{Highlighting} outperforms the \textit{Slider}.
\end{enumerate*}

We hypothesize that each interaction reduces error rates in comparison with the non-interactive interface, because both interactions comply with the \textit{rule of self-evidence} and address the \textit{adjacency task}. 
The \textit{rule of self-evidence} for multiple views prescribes the use  of ``perceptual cues to make relationships among multiple views more apparent to the user''~\cite{Baldonado2000}. The \textit{Highlighting} complies with this rule, by drawing attention to different instances of the same node across different time slices; the \textit{Slider}also complies with this rule, by allowing the user to select the maximum stability and fix node positions across different time slices.  
The \textit{adjacency task} (i.e., ``Given a node, find its adjacent nodes'') has been identified as the only graph-specific task~\cite{Lee2006}. The \textit{Highlighting} obviously addresses this task, as well as the \textit{Slider} does, by allowing the user to select the minimum stability and exploit the proximity \textit{Gestalt} principle~\cite{Bennett2007}.  
%
Conversely, we hypothesize that both interactions increase the task completion time in comparison with the non-interactive interface. We make this hypothesis in analogy with the existing comparative evaluations between animation and (static) timeline views~\cite{Farrugia2011}\cite{Archambault2011}, while we consider interactive timeline views as a middle way. More specifically, in terms of interaction costs~\cite{Lam2008}, the \textit{Highlighting} might increase the completion time because of the physical-motion cost of tracking elements with the mouse, while the \textit{Slider} might imply view-change costs of reinterpreting the perception when the layout rearranges. For both techniques, there might be the decision costs of forming goals, such as deciding whether the available interaction is useful to solve the given task, and how. Moreover, the simple fact that the GUI provides an interactive option might lead users to explore its use, in order to form a solving strategy before solving a task, or to possibly increase the confidence about the solution afterwards.  
Furthermore, we hypothesize that the \textit{Highlighting} will have better performance than the \textit{Slider}. We derive this hypothesis from the observation that the \textit{Highlighting} is a common and relatively simple interaction, which at least partially exploits pre-attentive processing, while the \textit{Slider} is based on a novel and complex concept. 
In other words, while the \textit{Highlighting} directly addresses the issue of connecting entities along two dimensions (time and graph structure), the \textit{Slider} implicitly introduces another dimension, since the stability lies in the parameter space of the layout algorithm.

\section{Analysis}
We preprocessed data collected from $64$ subjects in order to assess whether they were eligible for analysis and we had to discard one subject whose logs were corrupted. The analysis was then performed on data from $63$ subjects, consisting of $3024$ samples in total. 
We checked the completion times for normality with the Shapiro-Wilk goodness-of-fit test but the check failed. We then applied a logarithmic transformation to the completion times and checked again the normality with a positive result. The verification of the Gaussian condition assured the applicability of parametric tests; we could perform the analysis of variance through an ANOVA with the subject as a random variable. 
When the ANOVA found a factor to have a statistically significant effect, we compared the two levels of that factor with a pairwise post-hoc Student's t test; when the ANOVA found the interaction between factors to be statistically significant, we performed an all-pairs Tukey's  honestly significant difference (HSD) post-hoc test. 
The error can be understood as a dichotomous (i.e. binary) variable, since there are only two possible outcomes for each data sample (correct, not correct). Hence, we analysed the error by logistic regression as a generalized linear model (GLM) with a binomial distribution and a logit transformation as the link function, computing likelihood ratio statistics. When a factor was found to be a significant effect, we analysed the contrast between its levels in terms of pairwise comparisons between estimated marginal means.
%

\section{Results and discussion}

%

%
%
%

\begin{figure}[tbhp]
 \centering
 \subcaptionbox*{}
   {\includegraphics[width=0.49\columnwidth]{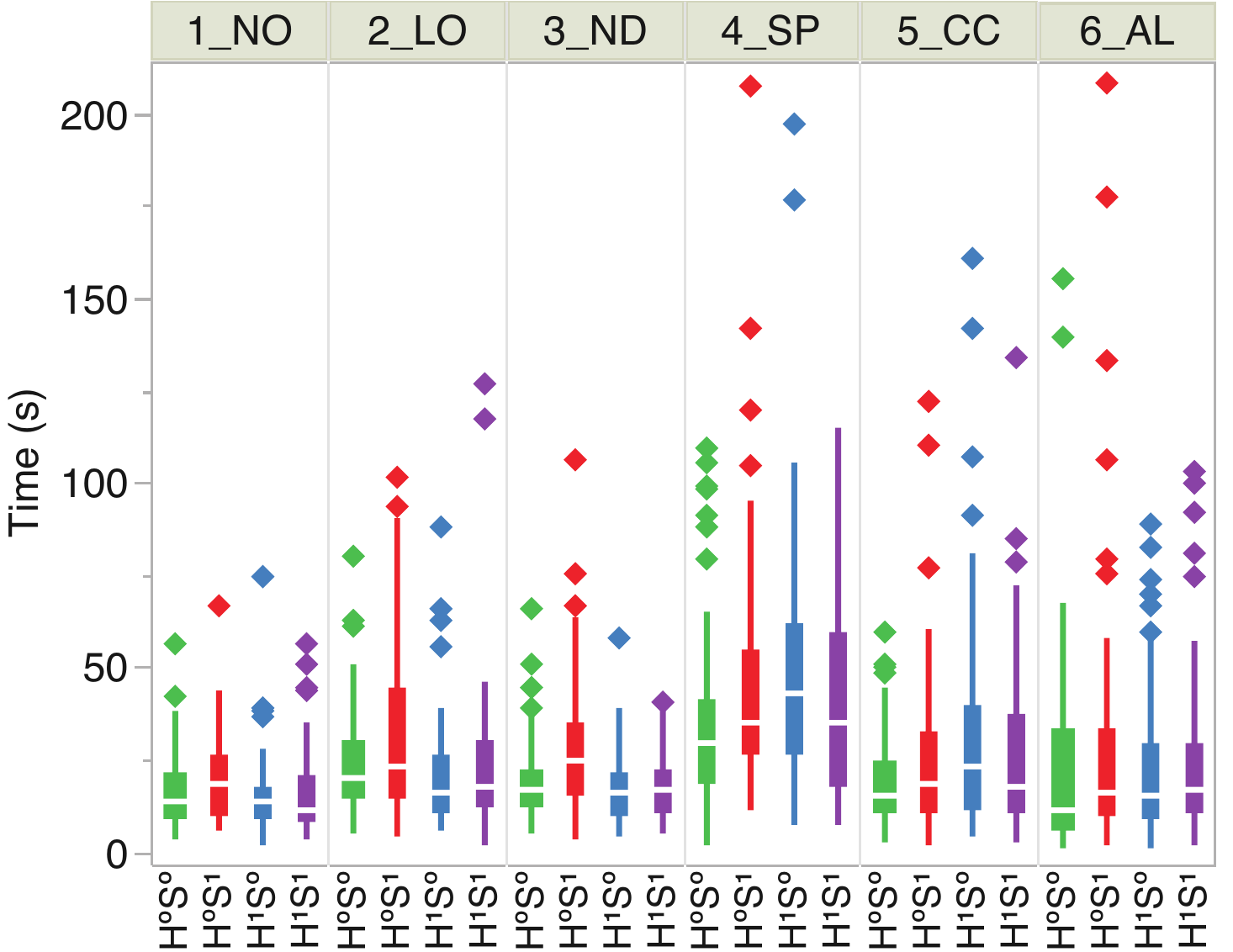}\label{fig:mean:time}}
 \subcaptionbox*{} 
   {\includegraphics[width=0.49\columnwidth]{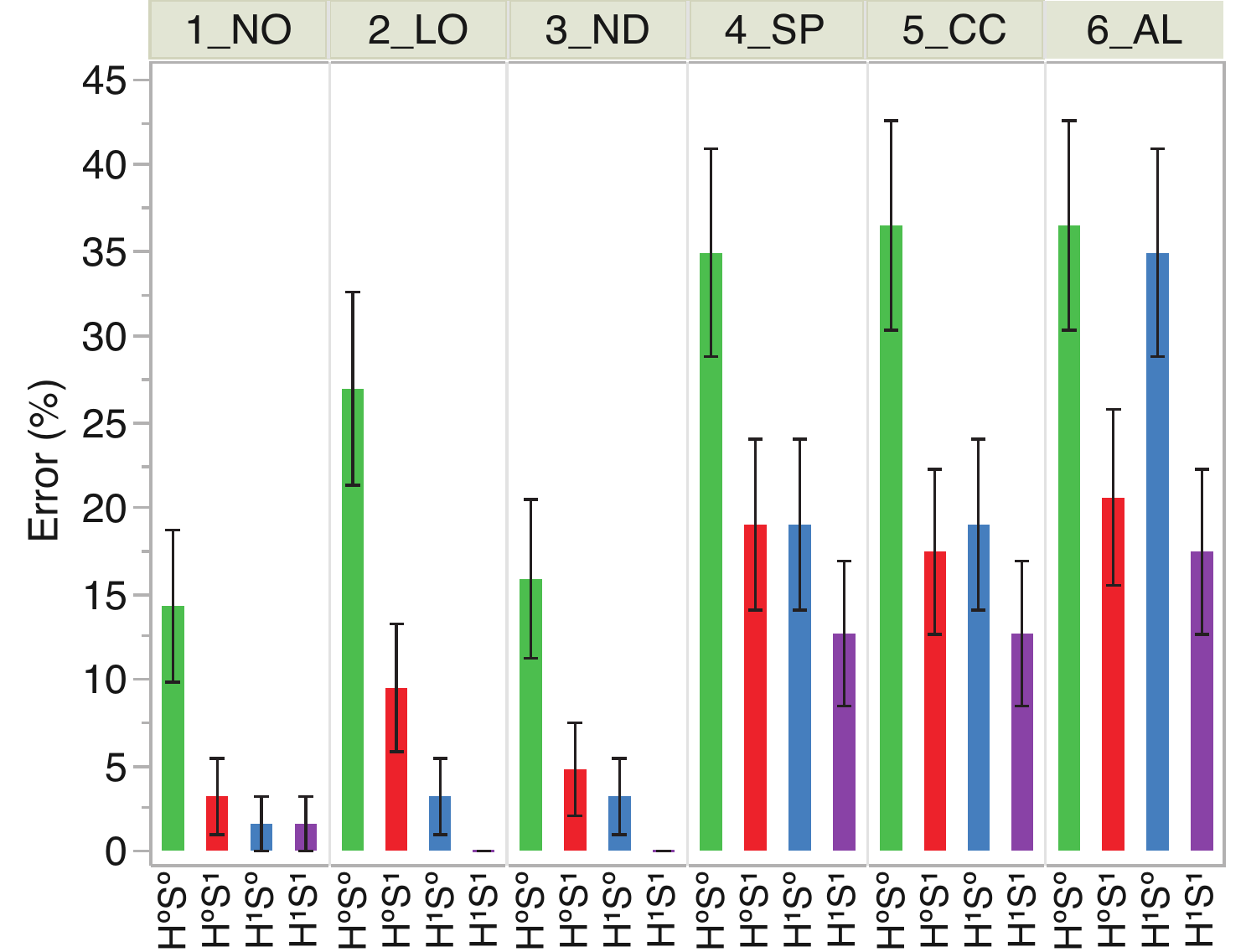}\label{fig:mean:error}}
 \caption{Time (left-hand side, as box plots) and error (right-hand side, as bars representing means and error bars representing standard error) by \textit{Highlighting} and \textit{Slider}, grouped by \textit{Task}. 
\textcolor{mygreen}{$\blacksquare$} $H^0S^0$
\textcolor{myred}{$\blacksquare$} $H^0S^1$
\textcolor{myblue}{$\blacksquare$} $H^1S^0$
\textcolor{mypurple}{$\blacksquare$} $H^1S^1$
}
 \label{fig:mean}
\end{figure}
%
%
%
%
\begin{figure*}[tbh]
 \centering

 
\begin{tabular*}{0.85\textwidth}{@{\extracolsep{\fill} } cccc }
   \quad Time & Error & Time & Error  \\
\end{tabular*}

  \subcaptionbox	*{Task 1.NO}{
   		{\includegraphics[width=0.232\columnwidth]{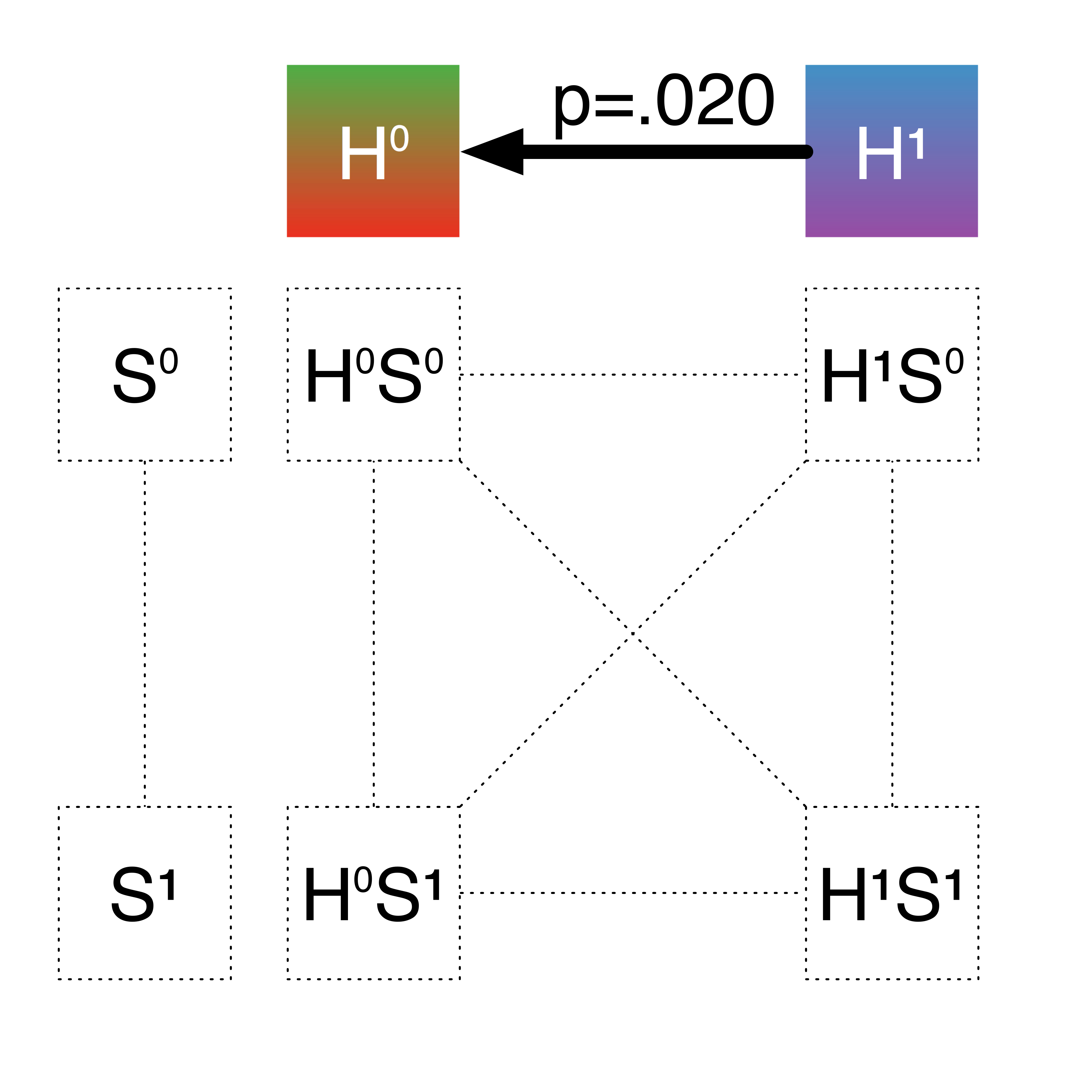}}
   		{\includegraphics[width=0.232\columnwidth]{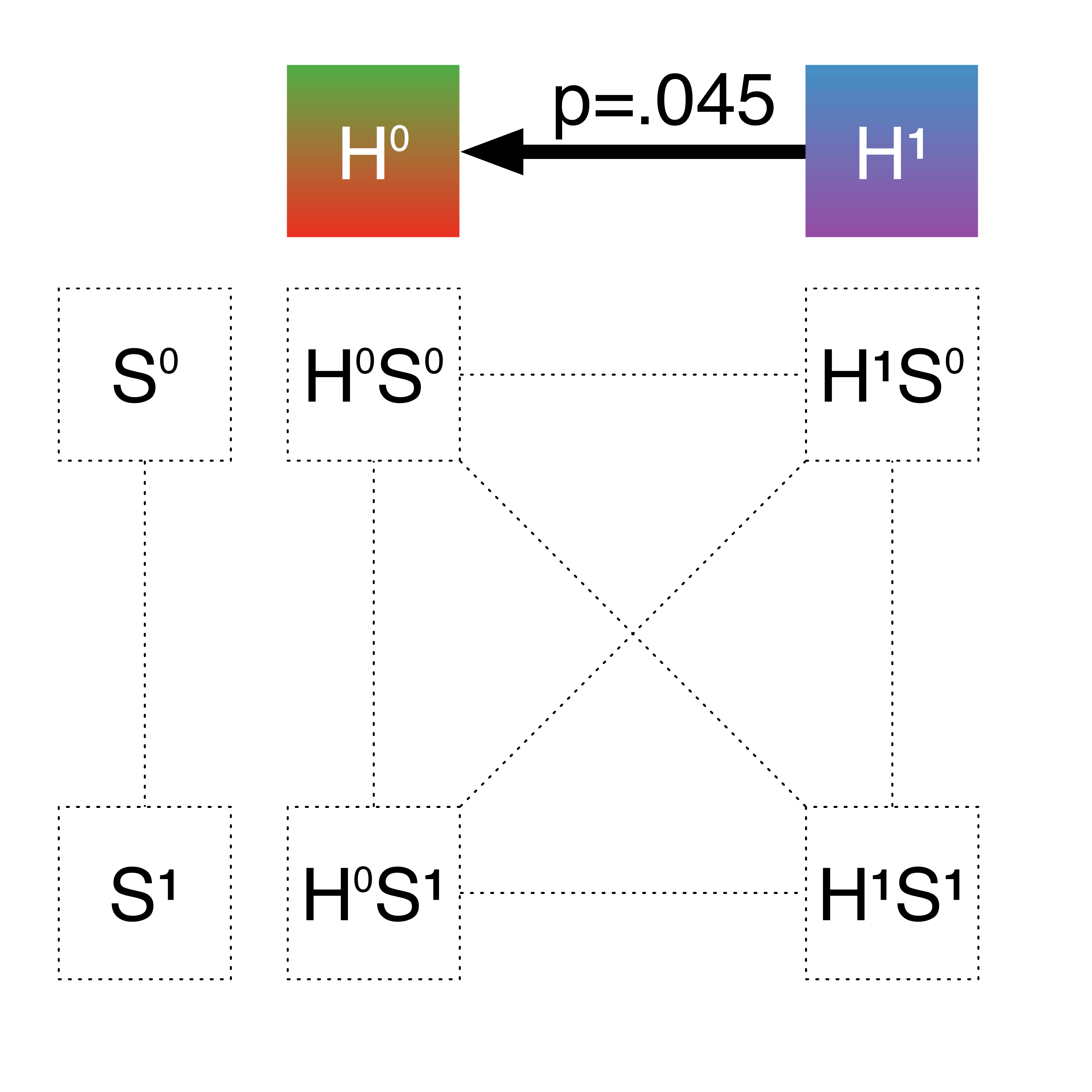}}
 }
 \subcaptionbox*	{Task 2.LO}{
   		{\includegraphics[width=0.232\columnwidth]{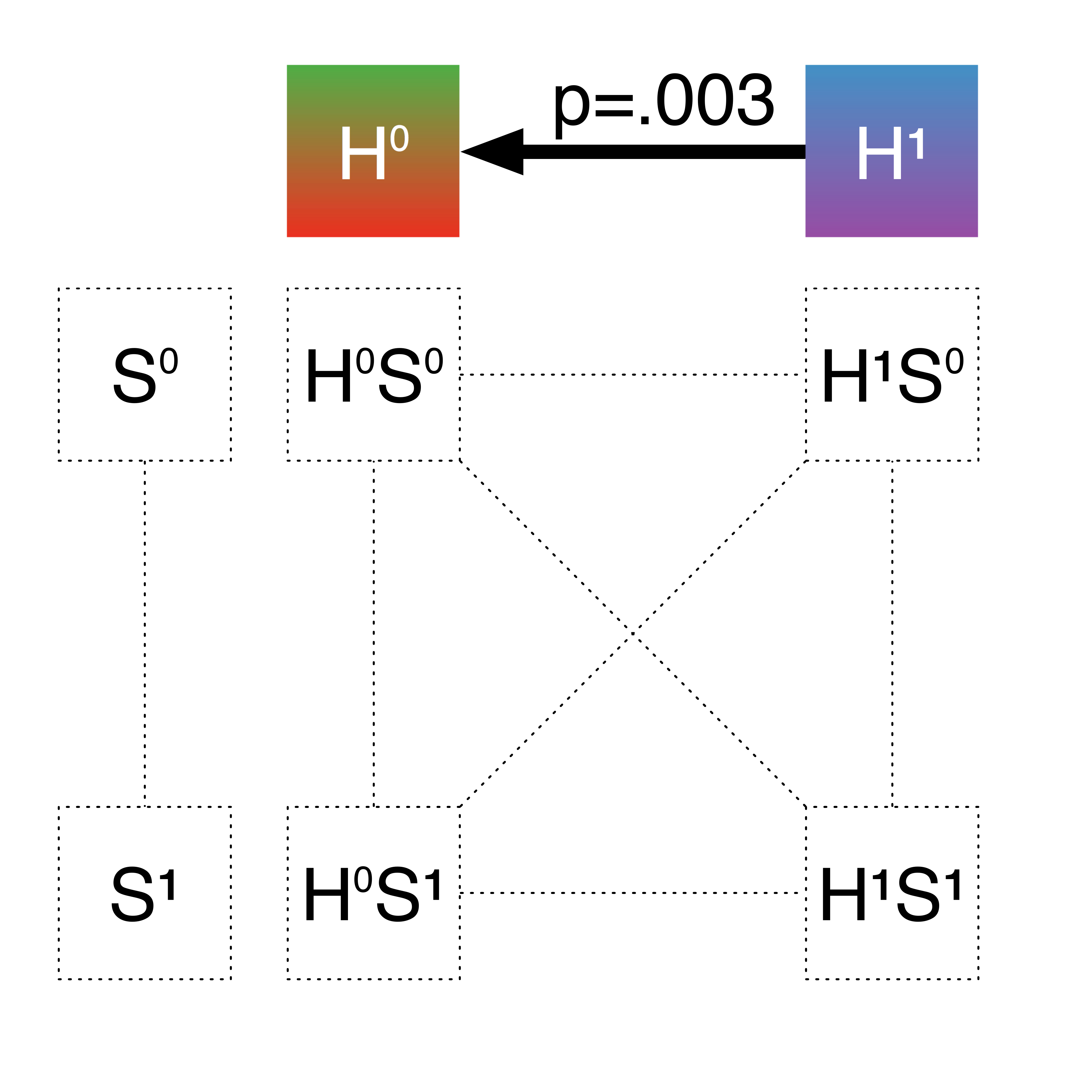}}
   		{\includegraphics[width=0.232\columnwidth]{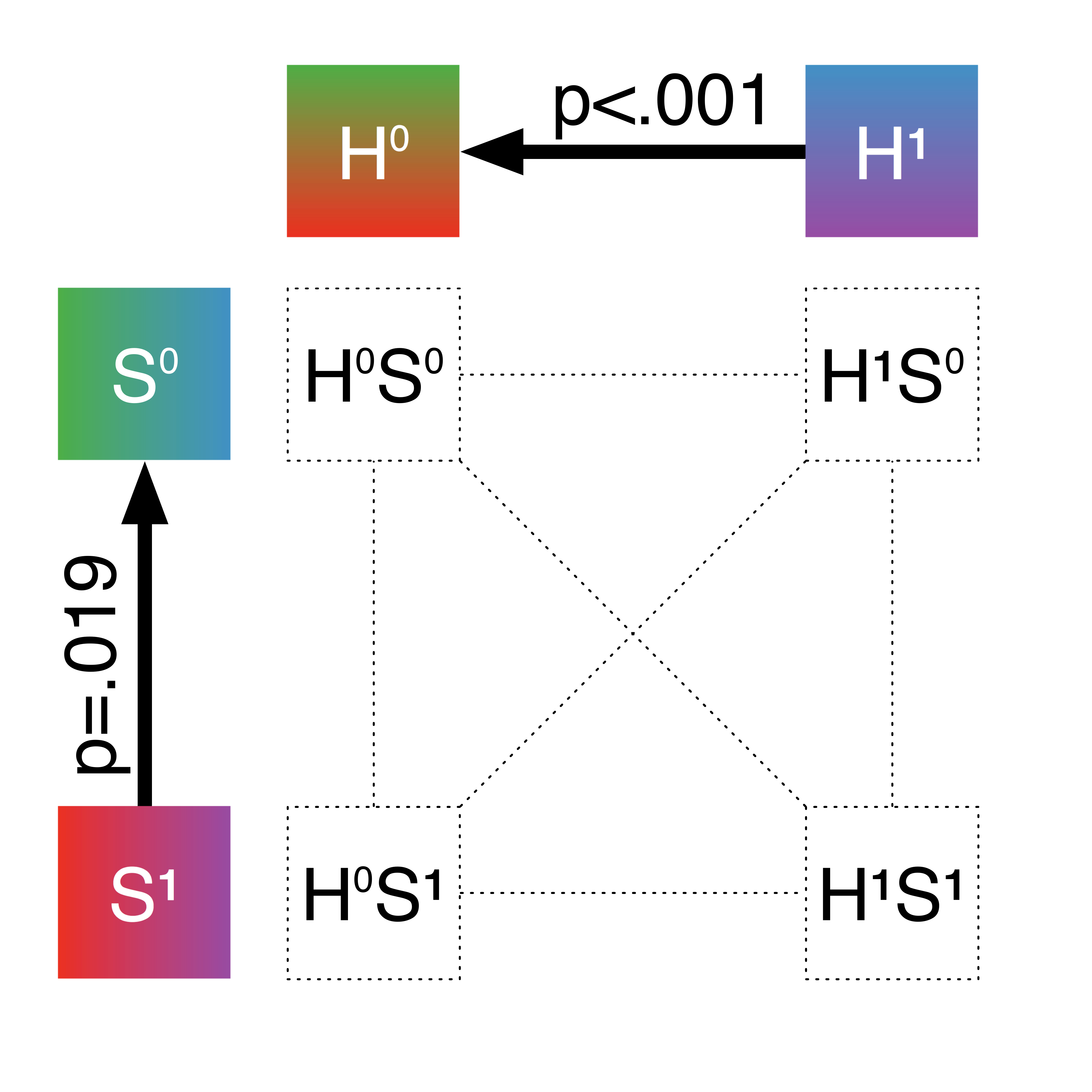}}
 }
  \subcaptionbox*	{Task 3.ND}{
   		{\includegraphics[width=0.232\columnwidth]{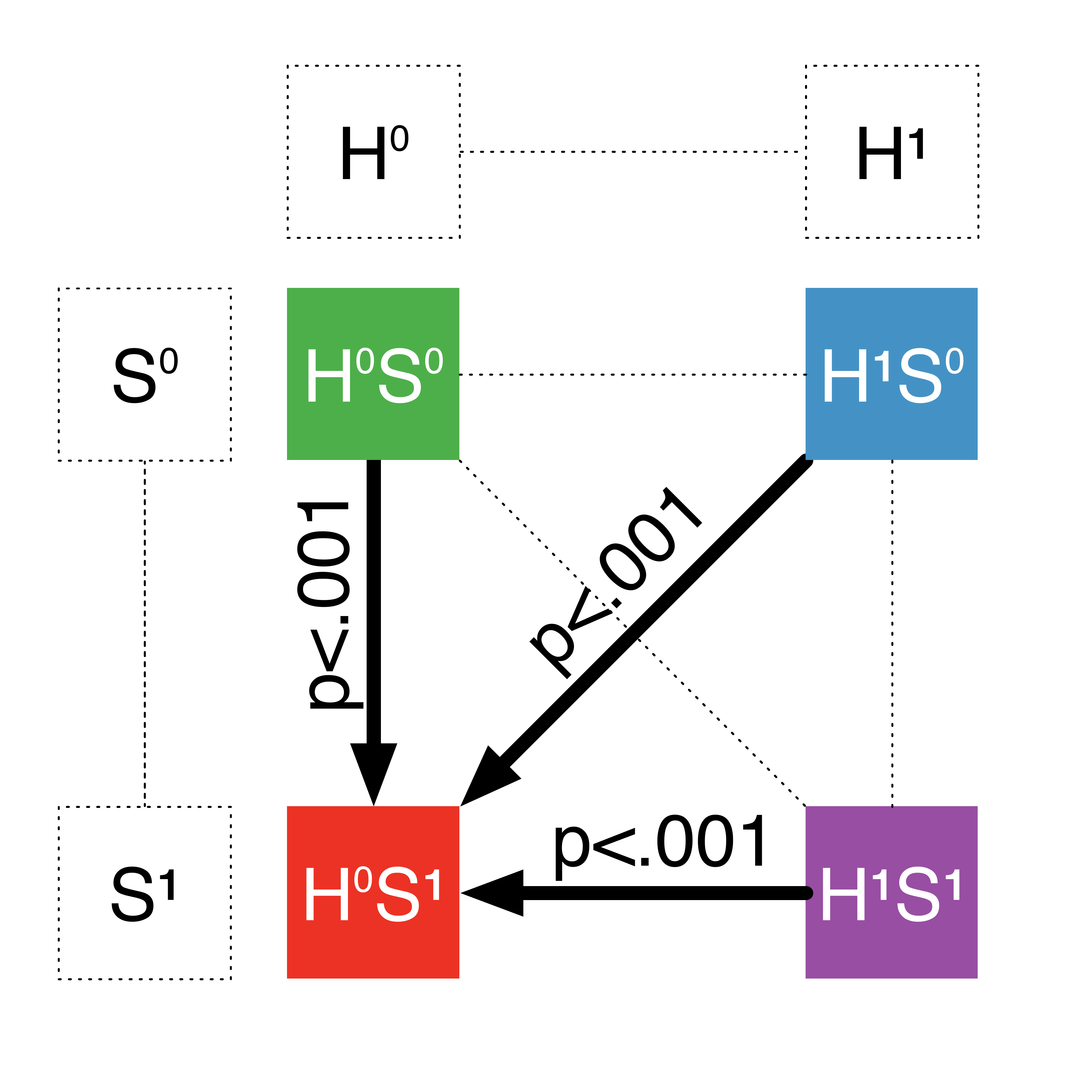}}
   		{\includegraphics[width=0.232\columnwidth]{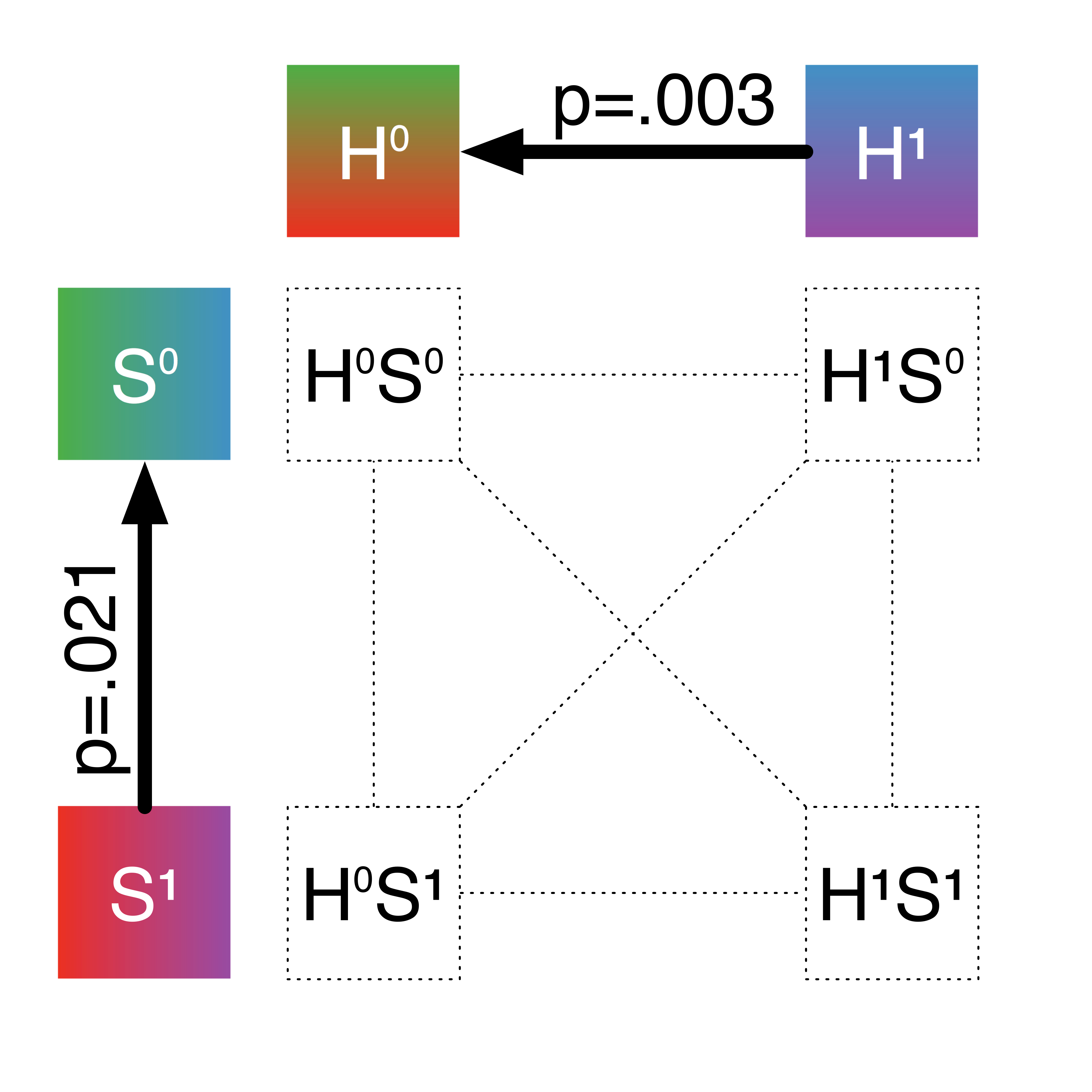}}
 }
 \subcaptionbox*{Task 4.SP}{
   		{\includegraphics[width=0.232\columnwidth]{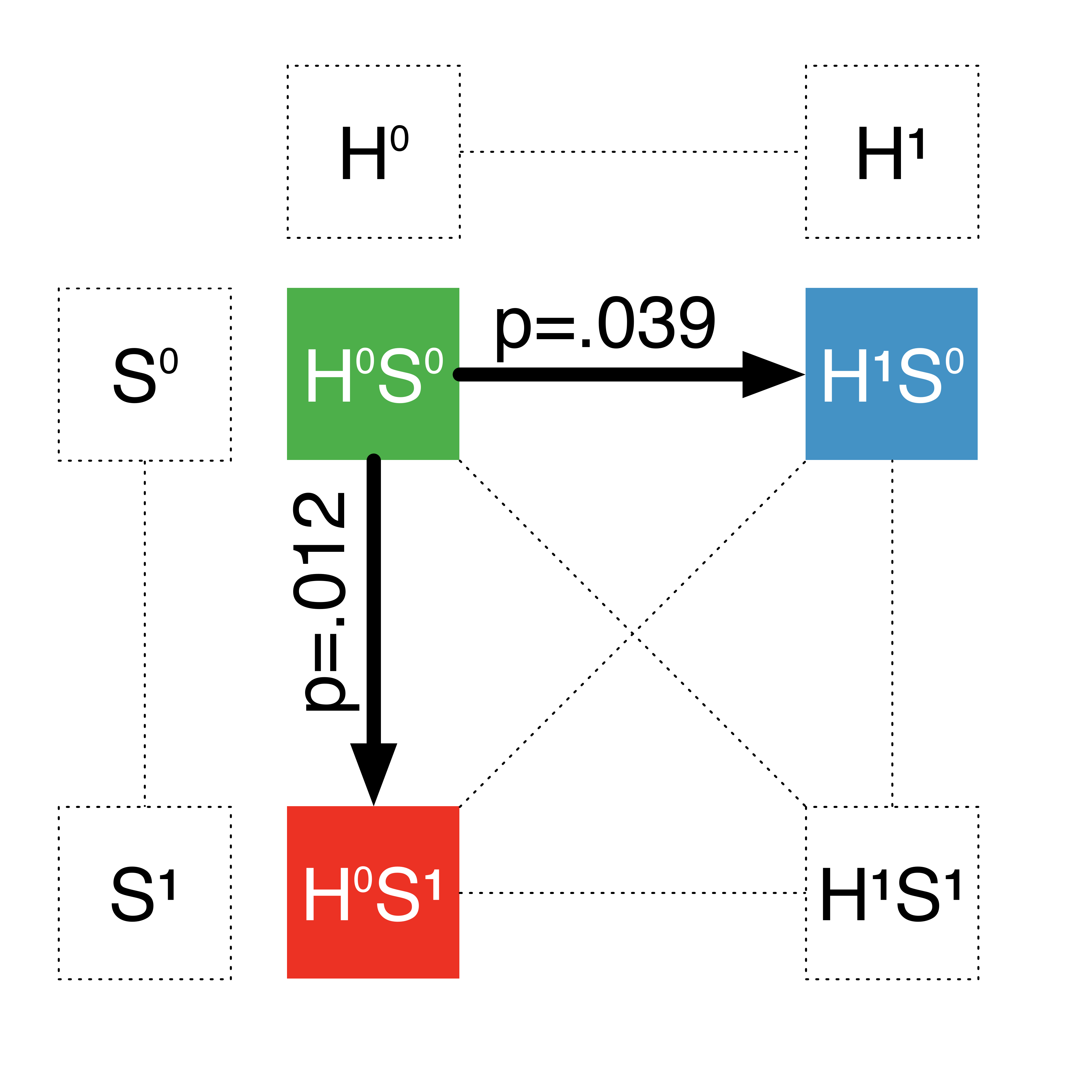}}
   		{\includegraphics[width=0.232\columnwidth]{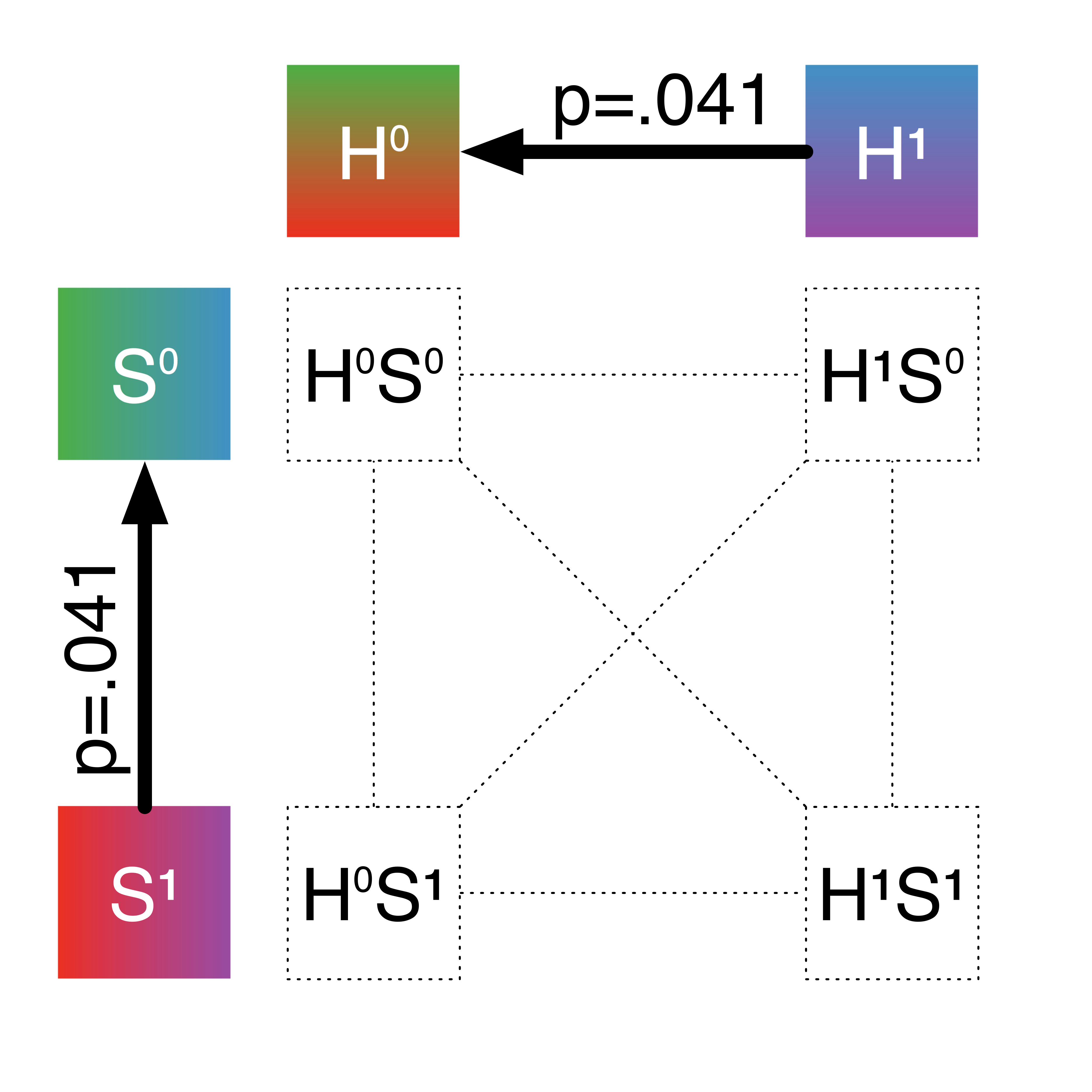}}
 }
  \subcaptionbox*	{Task 5.CC}{
   		{\includegraphics[width=0.232\columnwidth]{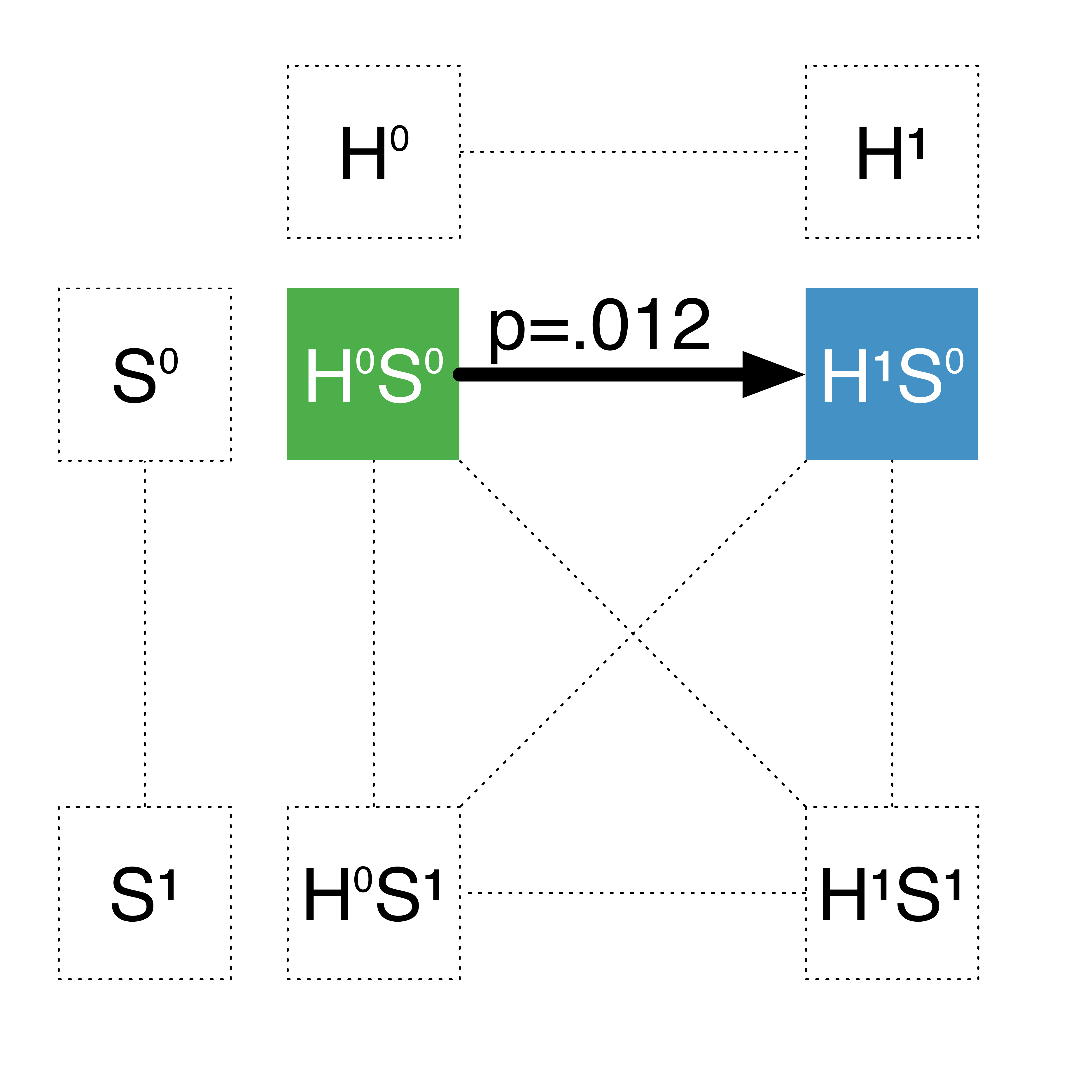}}
   		{\includegraphics[width=0.232\columnwidth]{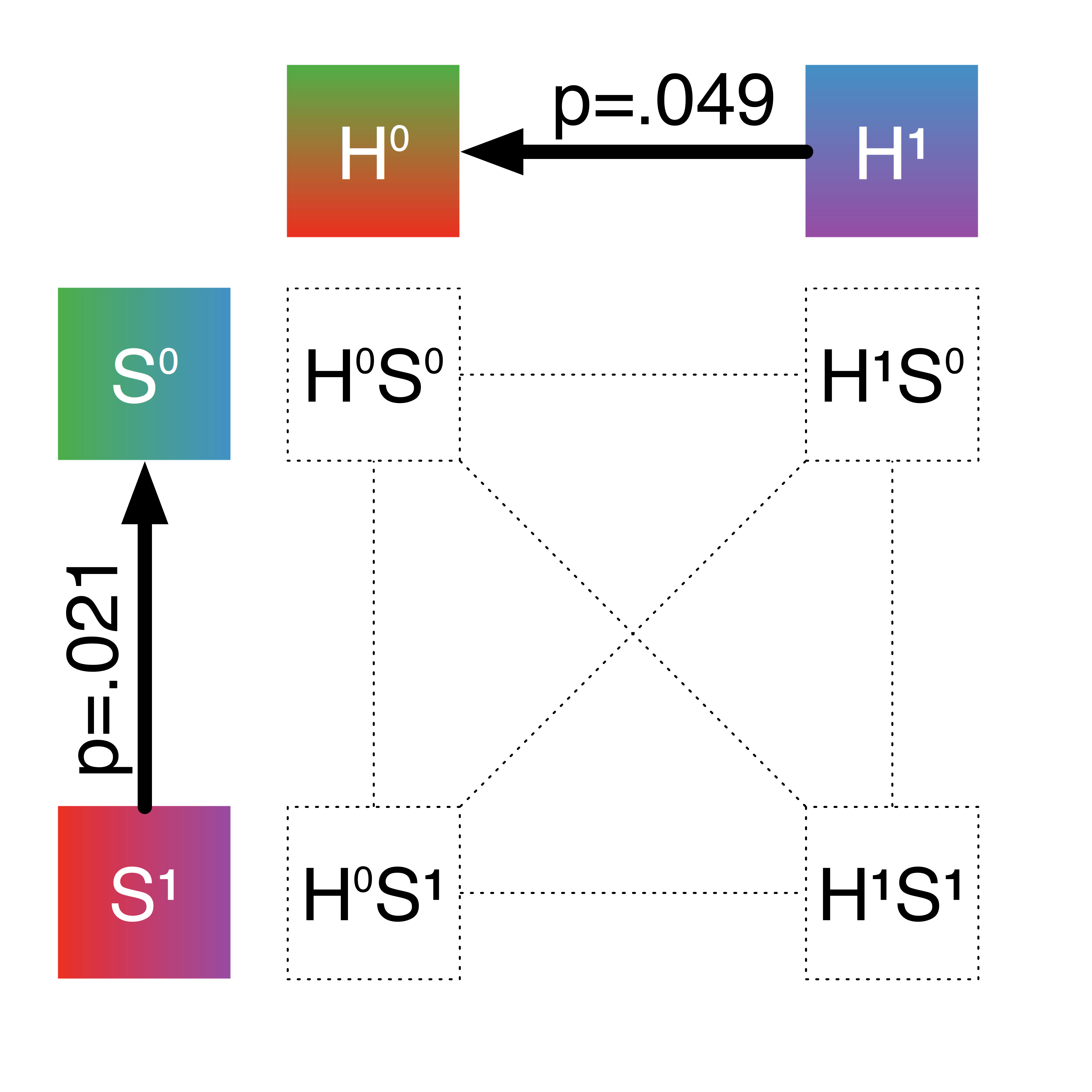}}
 }
 \subcaptionbox*{Task 6.AL}{
   		{\includegraphics[width=0.232\columnwidth]{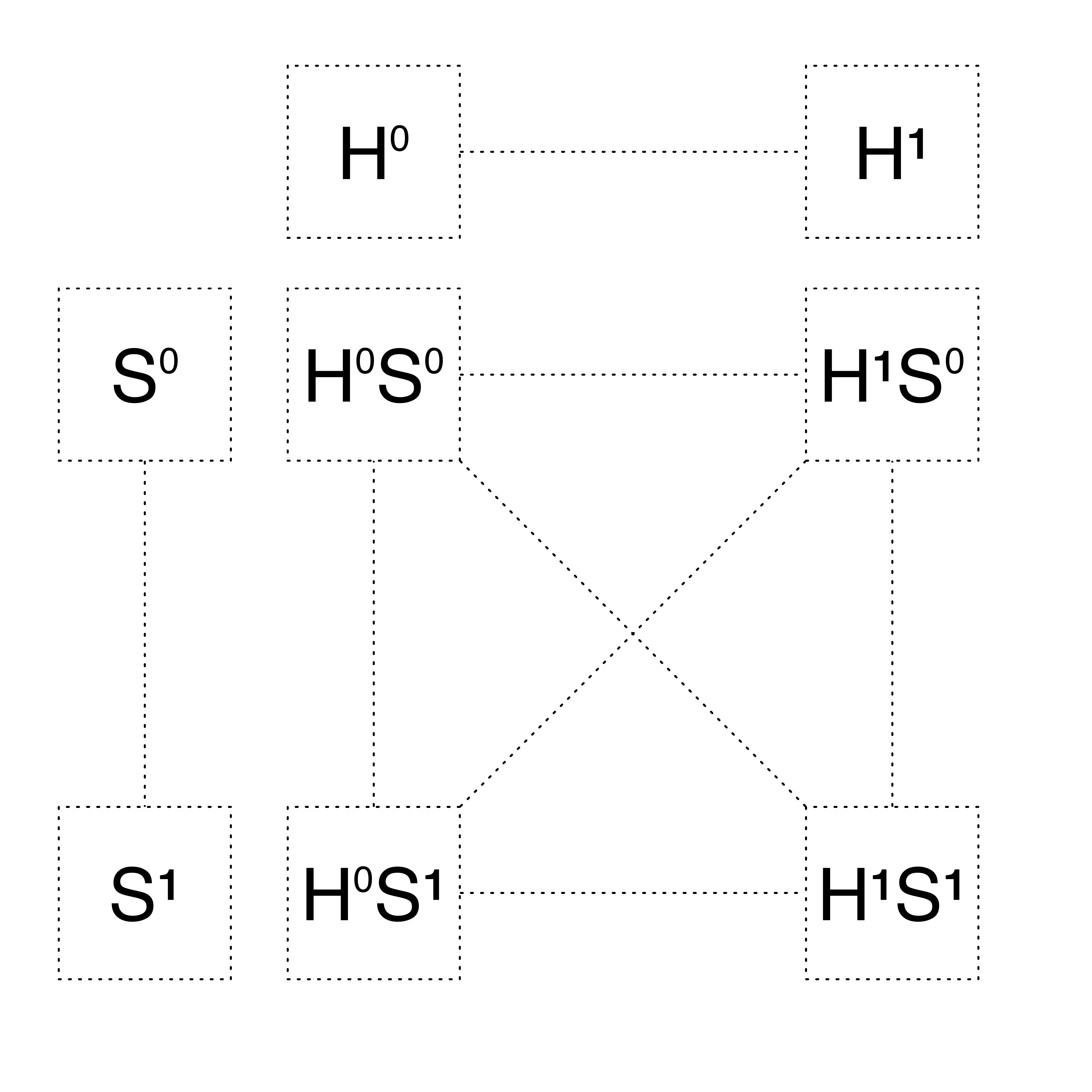}}
   		{\includegraphics[width=0.232\columnwidth]{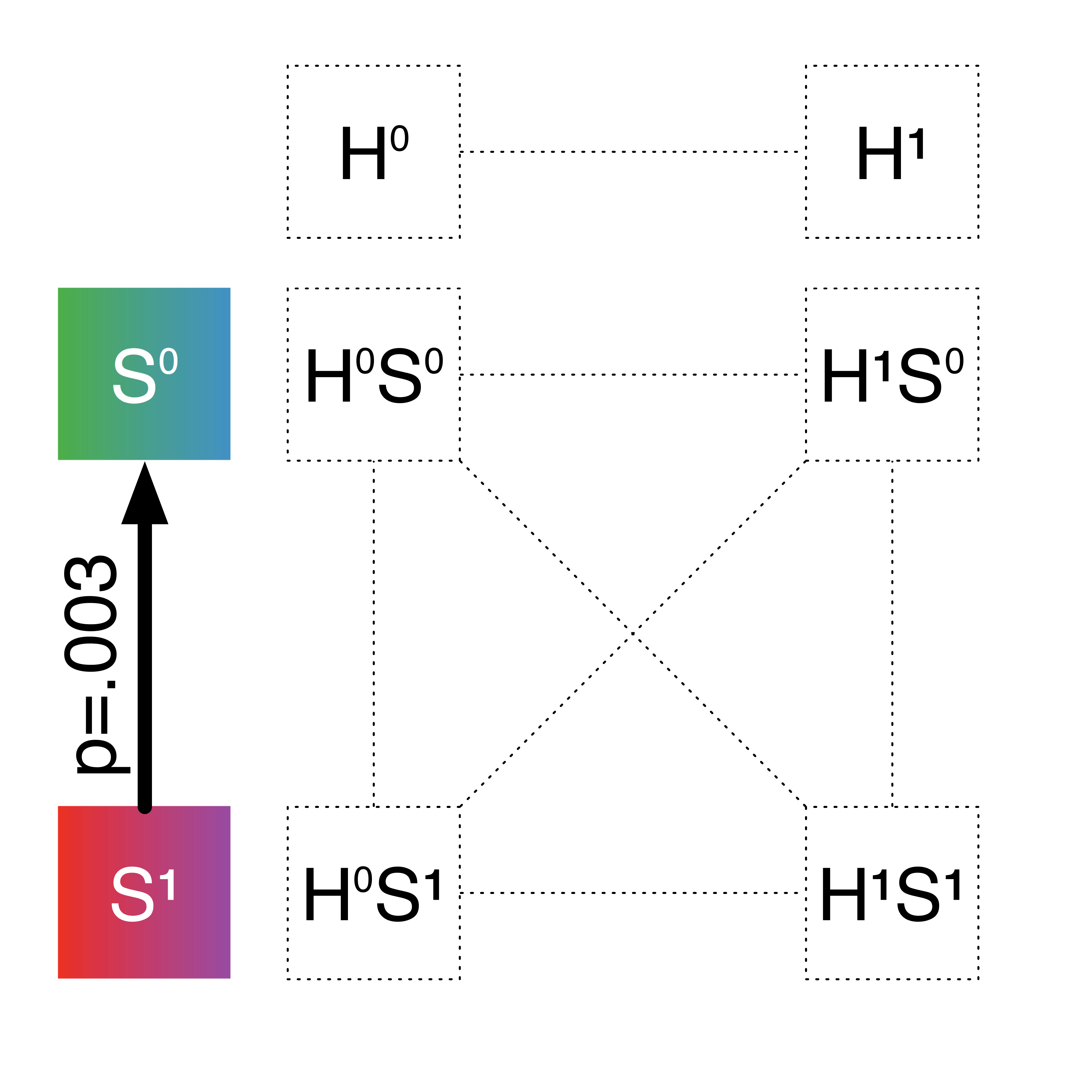}}
 }
 
 \caption{Statistically significant differences for time and error, by \textit{Task}. An arrow means that the source is faster or, respectively, more accurate than the destination, with the reported probability. 
Lines represent all-pairs comparisons between factor combinations (\textcolor{mygreen}{$\blacksquare$} $H^0S^0$, \textcolor{myred}{$\blacksquare$} $H^0S^1$, \textcolor{myblue}{$\blacksquare$} $H^1S^0$, and \textcolor{mypurple}{$\blacksquare$} $H^1S^1$), as well as pairwise comparisons by \textit{Highlighting} ($H^0$--$H^1$, top) and \textit{Slider} ($S^0$--$S^1$, left).
}
 \label{fig:arrowsTbyT}
 \end{figure*}
%

Figure~\ref{fig:mean} shows time and error by \textit{Highlighting} and \textit{Slider}, grouped by \textit{Task}; time is represented by box-plots with first, second (median), and third quartile, while error is represented by bars (mean) and error bars (standard error). Figure~\ref{fig:arrowsTbyT} shows statistically significant differences.  
In light of these results, we can verify our hypotheses.\\ 
\textbf{Hypothesis A} is partially confirmed. The \textit{Slider} decreases the error rate for all tasks but the easiest one (1.NO), and it increases the completion time for tasks 3.ND and 4.SP only. \\
\textbf{Hypothesis B} is partially confirmed. The \textit{Highlighting} decreases the error rate for all tasks but the most difficult one (6.AL); it increases completion times for tasks 4.SP and 5.CC, but it reduces it for task 2.LO, and does not affect the remaining tasks. \\
\textbf{Hypothesis C} is partially confirmed. The \textit{Highlighting} outperforms the \textit{Slider} for tasks 1.NO, 2.LO, and 3.ND. For task 4.SP, the \textit{Highlighting} and the \textit{Slider} score equally: each of them decreases the error rate (by the same amount) and also increases the completion time if used alone, but when used together they do not increase the completion time, showing a desirable effect interaction. 
For task 5.CC, both factors reduce the error rate, but the \textit{Highlighting} also increases the completion time when used alone. For task 6.AL, the only significant effect is that the \textit{Slider} reduces the error rate. 
%
%
%

Besides the verification of our hypothesis, which are mostly confirmed, our user study provides interesting insights about the differences between tasks. First of all, we observe that the differences in error rate and completion time among the tasks are significant, hence we can confirm that in general our tasks have different levels of difficulty. Secondly, we observe that the effectiveness of the tested interaction techniques varies with the levels of difficulty of the tasks. In a very brief but accurate summary we can say that, for easier tasks, the \textit{Highlighting} decreases error rates and in some cases even decreases completion times; conversely, for more difficult tasks, it is the \textit{Slider} that decreases error rates. Moreover, for tasks 3.ND, 4.SP and 5.CC, one technique increases completion times if used alone, but it does not if used in combination with the other one. Looking back at the classification of our tasks (Table~\ref{tab:tasks}), we can also identify the relevant aspects. We can observe that, for those tasks involving simpler temporal features of distinguishable single entities (1.NO and 2.LO), or indistinguishable groups (3.ND), the \textit{Highlighting} is more effective. For those tasks that refer to more complex temporal features at the graph level, even if indistinguishable (5.CC and 6.AL), the \textit{Slider} is more effective. Task 4.SP is about the changes of the geodesic distance between two nodes and requires the distinct identification of several nodes and links along the shortest path. In this case both techniques are equally accurate; if (and only if) they are used together, they do not even slow down the analysis. We can conjecture (by also considering our observations during the pilot experiments) that during the completion of a such complex task, the \textit{Slider} can be used to switch back and forth between the minimum stability (to guess geodesic distances and shortest paths based on Euclidean distances) and the maximum stability (to identify instances of nodes and links across different time slices), while the \textit{Highlighting} helps with tracking objects. 
As for task 6.AL, the \textit{Slider} resulted to be effective; we hypothesize (by also considering our pilot observations) that subjects simply set the minimum stability and looked at the total graph area as an estimator of the density. We would have expected the \textit{Highlighting} to be also effective, since the analysis of the degree a few central nodes might provide a good estimator of the graph density, given the power-law distribution of real-world networks. The results show that the test subjects did not exploit this expert strategy. 
\subsubsection{\textit{Design implications}} 
Both the \textit{Slider} and the \textit{Highlighting} are effective interaction techniques for dynamic graph visualization, and their use generally improves user performances. In those circumstances where it might be not possible to include them both (for example, if the color channel is employed to encode attributes of multivariate graphs, or if the GUI is already overloaded with many controls), our evaluation provides an indication to designers according to the user tasks to be supported. Our results suggest that the \textit{Highlighting} is indicated for tasks involving temporal features of distinguishable single entities or indistinguishable groups, while the \textit{Slider} is indicated for tasks involving complex temporal behaviours at the graph level. The joint use of both interactions is beneficial for the most complex task involving temporal behaviours of connectivity paths. 
\section{Conclusion}
We have presented an evaluation of two interaction techniques for dynamic graph visualization, namely the interactive control of the layout stability and the interactive highlighting of adjacent nodes and links. The results mostly confirm our hypotheses: both interactions decrease the error rate, in some cases at the cost of a longer completion time. We observed significant differences between tasks, with the highlighting performing better for some tasks, and the stability control performing better for others.  
We acknowledge the limitations of our experiment, whose findings might not be directly generalizable to large-scale datasets. The highlighting interaction for dynamic graphs is much more complex then the standard connectivity highlighting for static graphs, and may require training to be understood and used effectively. The stability control might have a different effect when combined with 3D visualization and interaction techniques (e.g., the vertigo zoom~\cite{Federico2012}). However, our study provides preliminary clues for visualization designers who need to choose the most appropriate interaction technique for their users' tasks. Further studies are needed to obtain a comprehensive understanding of the role of interaction in visualization of dynamic graphs.

%
%

\subsubsection*{\textbf{Acknowledgements}} 
The authors wish to thank the anonymous study subjects for their participation, as well as Theresia Gschwandtner, Simone Kriglstein, and Margit Pohl for their feedback on the manuscript.\\
This work was partially supported by the Austrian Research Promotion Agency (FFG), project \textit{Expand}, grant 835937.

\nocite{Kerren2014}
\bibliographystyle{splncs03}
\bibliography{GD2016}

\begin{thebibliography}{10}
\providecommand{\url}[1]{\texttt{#1}}
\providecommand{\urlprefix}{URL }

\bibitem{Ahn2014}
Ahn, J.W., Plaisant, C., Shneiderman, B.: A task taxonomy for network evolution
  analysis. Visualization and Computer Graphics, IEEE Trans.  20(3),  365--376
  (Mar 2014)

\bibitem{Aigner2013}
Aigner, W., Hoffmann, S., Rind, A.: Evalbench: A software library for
  visualization evaluation. Computer Graphics Forum  32(3pt1),  41--50 (2013)

\bibitem{Archambault2014}
Archambault, D., Abello, J., Kennedy, J., Kobourov, S., Ma, K.L., Miksch, S.,
  Muelder, C., Telea, A.: {Temporal Multivariate Networks}. In: Kerren et~al.
  \cite{Kerren2014}, pp. 151--174

\bibitem{Archambault2011}
Archambault, D., Purchase, H., Pinaud, B.: Animation, small multiples, and the
  effect of mental map preservation in dynamic graphs. Visualization and
  Computer Graphics, IEEE Trans.  17(4),  539--552 (Apr 2011)

\bibitem{Archambault2012}
Archambault, D., Purchase, H.C.: The mental map and memorability in dynamic
  graphs. In: Proc. Pacific Visualization Symp. pp. 89--96. PacificVis12, IEEE,
  Washington, DC (2012)

\bibitem{Archambault2013}
Archambault, D., Purchase, H.C.: The map in the mental map: Experimental
  results in dynamic graph drawing. Intern. Journal of Human-Computer Studies
  71(11),  1044 -- 1055 (2013)

\bibitem{Archambault2014b}
Archambault, D., Purchase, H.C.: On the application of experimental results in
  dynamic graph drawing. In: Proc. 1st Int. Workshop on Graph Visualization in
  Practice. GViP14, vol. 1244, pp. 73--77. CEUR-WS (2014)

\bibitem{Bach2014}
Bach, B., Pietriga, E., Fekete, J.D.: {GraphDiaries: Animated Transitions and
  Temporal Navigation for Dynamic Networks}. Visualization and Computer
  Graphics, IEEE Trans.  20(5),  740--754 (May 2014)

\bibitem{Baldonado2000}
Baldonado, M.Q.W., Woodruff, A., Kuchinsky, A.: Guidelines for using multiple
  views in information visualization. In: Proceedings of the Working Conference
  on Advanced Visual Interfaces. pp. 110--119. AVI '00, ACM, New York, NY, USA
  (2000)

\bibitem{Beck2014}
Beck, F., Burch, M., Diehl, S., Weiskopf, D.: {The State of the Art in
  Visualizing Dynamic Graphs}. In: Borgo, R., Maciejewski, R., Viola, I. (eds.)
  EuroVis - STARs. pp. 83--103. Eurographics Association (2014)

\bibitem{Becker1987}
Becker, R.A., Cleveland, W.S.: Brushing scatterplots. Technometrics  29(2),
  127--142 (1987)

\bibitem{Bennett2007}
Bennett, C., Ryall, J., Spalteholz, L., Gooch, A.: {The Aesthetics of Graph
  Visualization}. In: Cunningham, D.W., Meyer, G., Neumann, L. (eds.)
  Computational Aesthetics in Graphics, Visualization, and Imaging. The
  Eurographics Association (2007)

\bibitem{Boyandin2012}
Boyandin, I., Bertini, E., Lalanne, D.: A qualitative study on the exploration
  of temporal changes in flow maps with animation and small-multiples. Computer
  Graphics Forum  31(3pt2),  1005--1014 (2012)

\bibitem{Brandes2012}
Brandes, U., Mader, M.: A quantitative comparison of stress-minimization
  approaches for offline dynamic graph drawing. In: van Kreveld, M., Speckmann,
  B. (eds.) Graph Drawing, LNCS, vol. 7034, pp. 99--110. Springer (2012)

\bibitem{Card1999}
Card, S.K., Mackinlay, J.D., Shneiderman, B. (eds.): Readings in Information
  Visualization: Using Vision to Think. Morgan Kaufmann, San Francisco, CA
  (1999)

\bibitem{Coleman1996}
Coleman, M.K., Parker, D.S.: Aesthetics-based graph layout for human
  consumption. Software: Practice and Experience  26(12),  1415--1438 (1996)

\bibitem{Farrugia2011}
Farrugia, M., Quigley, A.: Effective temporal graph layout: A comparative study
  of animation versus static display methods. Information Visualization  10(1),
   47--64 (2011)

\bibitem{Federico2012}
Federico, P., Aigner, W., Miksch, S., Windhager, F., Smuc, M.: Vertigo zoom:
  Combining relational and temporal perspectives on dynamic networks. In:
  Proceedings of the International Working Conference on Advanced Visual
  Interfaces. pp. 437--440. AVI '12, ACM, New York, NY, USA (2012)

\bibitem{Federico2011}
Federico, P., Aigner, W., Miksch, S., Windhager, F., Zenk, L.: A visual
  analytics approach to dynamic social networks. In: Proc. Intern. Conf. on
  Knowledge Management and Knowledge Technologies. pp. 47:1--47:8. i-KNOW '11,
  ACM, New York, NY (2011)

\bibitem{Ghani2012}
Ghani, S., Elmqvist, N., Yi, J.S.: Perception of animated node-link diagrams
  for dynamic graphs. Computer Graphics Forum  31(3pt3),  1205--1214 (2012)

\bibitem{Heer2005}
Heer, J., Boyd, D.: Vizster: visualizing online social networks. In: IEEE
  Symposium on Information Visualization, 2005. INFOVIS 2005. pp. 32--39 (Oct
  2005)

\bibitem{Heer2005a}
Heer, J., Card, S.K., Landay, J.A.: Prefuse: A toolkit for interactive
  information visualization. In: Proceedings of the SIGCHI Conference on Human
  Factors in Computing Systems. pp. 421--430. CHI '05, ACM, New York, NY, USA
  (2005)

\bibitem{Herman2000}
Herman, I., Melan\c{c}on, G., Marshall, M.S.: Graph visualization and
  navigation in information visualization: A survey. Visualization and Computer
  Graphics, IEEE Trans.  6(1),  24--43 (Jan 2000)

\bibitem{Keim2002}
Keim, D.A.: Information visualization and visual data mining. IEEE Transactions
  on Visualization and Computer Graphics  8(1),  1--8 (Jan 2002)

\bibitem{Kerracher2014}
Kerracher, N., Kennedy, J., Chalmers, K.: {The Design Space of Temporal Graph
  Visualisation}. In: Elmqvist, N., Hlawitschka, M., Kennedy, J. (eds.) EuroVis
  - Short Papers. pp. 7--11. Eurographics Association (2014)

\bibitem{Kerren2014}
Kerren, A., Purchase, H., Ward, M. (eds.): Multivariate Network Visualization,
  LNCS, vol. 8380. Springer (2014)

\bibitem{Lam2008}
Lam, H.: A framework of interaction costs in information visualization.
  Visualization and Computer Graphics, IEEE Trans.  14(6),  1149--1156 (Nov
  2008)

\bibitem{Lee2006}
Lee, B., Plaisant, C., Parr, C.S., Fekete, J.D., Henry, N.: Task taxonomy for
  graph visualization. In: Proc. AVI Workshop on Beyond Time and Errors: Novel
  Evaluation Methods for Information Visualization. pp. 1--5. BELIV '06, ACM,
  New York, NY (2006)

\bibitem{McGuffin2009}
McGuffin, M., Jurisica, I.: Interaction techniques for selecting and
  manipulating subgraphs in network visualizations. Visualization and Computer
  Graphics, IEEE Trans.  15(6),  937--944 (Nov 2009)

\bibitem{Misue1995}
Misue, K., Eades, P., Lai, W., Sugiyama, K.: Layout adjustment and the mental
  map. Journal of Visual Languages \& Computing  6(2),  183 -- 210 (1995)

\bibitem{Moscovich2009}
Moscovich, T., Chevalier, F., Henry, N., Pietriga, E., Fekete, J.D.:
  Topology-aware navigation in large networks. In: Proc. SIGCHI Conf. on Human
  Factors in Computing Systems. pp. 2319--2328. CHI '09, ACM, New York, NY
  (2009)

\bibitem{Pike2009}
Pike, W.A., Stasko, J., Chang, R., O'Connell, T.A.: The science of interaction.
  Information Visualization  8(4),  263--274 (2009)

\bibitem{Pretorius2014}
Pretorius, A., Purchase, H., Stasko, J.: Tasks for multivariate network
  analysis. In: Kerren et~al.  \cite{Kerren2014}, pp. 77--95

\bibitem{Purchase2007}
Purchase, H., Hoggan, E., G\"{o}rg, C.: How important is the mental map? {A}n
  empirical investigation of a dynamic graph layout algorithm. In: Kaufmann,
  M., Wagner, D. (eds.) Graph Drawing, LNCS, vol. 4372, pp. 184--195. Springer
  (2007)

\bibitem{Purchase2008}
Purchase, H., Samra, A.: Extremes are better: Investigating mental map
  preservation in dynamic graphs. In: Stapleton, G., Howse, J., Lee, J. (eds.)
  Diagrammatic Representation and Inference, LNCS, vol. 5223, pp. 60--73.
  Springer (2008)

\bibitem{Rey2010}
Rey, G.D., Diehl, S.: Controlling presentation speed, labels, and tooltips in
  interactive animations. Journal of Media Psychology: Theories, Methods, and
  Applications  22(4),  160 (2010)

\bibitem{Saffrey2008}
Saffrey, P., Purchase, H.: The mental map versus static aesthetic compromise in
  dynamic graphs: A user study. In: Proc. Conf. on Australasian User Interface.
  pp. 85--93. AUIC '08, Australian Comp. Soc. (2008)

\bibitem{Smuc2014}
Smuc, M., Federico, P., Windhager, F., Aigner, W., Zenk, L., Miksch, S.: How do
  you connect moving dots? insights from user studies on dynamic network
  visualizations. In: Huang, W. (ed.) Handbook of Human Centric Visualization,
  pp. 623--650. Springer New York (2014)

\bibitem{Spence2007}
Spence, R.: Information visualization : design for interaction. Addison Wesley,
  Harlow, England New York (2007)

\bibitem{VanDeBunt1999}
Van De~Bunt, G.G., Van~Duijn, M.A.J., Snijders, T.A.B.: Friendship networks
  through time: An actor-oriented dynamic statistical network model. Comput.
  Math. Organ. Theory  5(2),  167--192 (Jul 1999)

\bibitem{vonLandesberger2011}
{von Landesberger}, T., Kuijper, A., Schreck, T., Kohlhammer, J., van Wijk, J.,
  Fekete, J.D., Fellner, D.: Visual analysis of large graphs: State-of-the-art
  and future research challenges. Computer Graphics Forum  30(6),  1719--1749
  (2011)

\bibitem{Ware2004}
Ware, C.: Information visualization : perception for design. Morgan Kaufman,
  San Francisco, CA (2004)

\bibitem{Ware2005}
Ware, C., Bobrow, R.: Supporting visual queries on medium-sized node–link
  diagrams. Information Visualization  4(1),  49--58 (2005)

\bibitem{Wybrow2014}
Wybrow, M., Elmqvist, N., Fekete, J.D., von Landesberger, T., van Wijk, J.,
  Zimmer, B.: Interaction in the visualization of multivariate networks. In:
  Kerren et~al.  \cite{Kerren2014}, pp. 97--125

\bibitem{Yi2007}
Yi, J.S., ah~Kang, Y., Stasko, J., Jacko, J.: Toward a deeper understanding of
  the role of interaction in information visualization. Visualization and
  Computer Graphics, IEEE Trans.  13(6),  1224--1231 (Nov 2007)

\bibitem{Zaman2011}
Zaman, L., Kalra, A., Stuerzlinger, W.: The effect of animation, dual view,
  difference layers, and relative re-layout in hierarchical diagram
  differencing. In: Proc. Conf. Graphics Interface. pp. 183--190. GI '11,
  Canadian Human-Computer Comm. Soc. (2011)

\end{thebibliography}

\end{document}